\documentstyle[12pt]{article}
\newlength{\mytopmargin}
\newlength{\myleftmargin}
\begin{document}

\vspace{4cm}
\noindent
{\large
{\bf
A Testbench for the Nested Dipole Hypothesis
of Kosterlitz \\and Thouless}}
\vspace{5mm}

\noindent
A.Alastuey\footnote{email: alastuey@physique.ens-lyon.fr}\\
\noindent
Ecole Normale Sup\'{e}rieure de Lyon,Laboratoire de Physique,
Unit\'{e} de Recherche Associe\'{e} 1325 au Centre National de la Recherche
Scientifique,
46,all\'{e}e d'Italie,69364 Lyon Cedex 07,France
\vspace{5mm}

\noindent
P.J.Forrester\footnote{email: matpjf@maths.mu.oz.au}\\
\noindent
Department of Mathematics, University of Melbourne,Parkville,Victoria
3052,Australia
\vspace{2cm}

\small
\begin{quote}
We consider the two-dimensional one-component plasma without a background
and confined to a half plane near a metal wall. The particles are also
subjected to an external potential acting perpendicular to the wall with
an inverse power law Boltzmann factor.
The model has a known solvable isotherm which exhibits a
Kosterlitz-Thouless type transition from a conductive to an insulator phase
as the power law is varied. This allows predictions of theoretical methods
of analysing the Kosterlitz-Thouless transition to be compared with the
exact solution. In particular we calculate
the asymptotic density   profile  by resumming its low fugacity
expansion near the zero-density critical coupling
in the insulator phase, and solving a mean-field equation deduced from the
first BGY equation. Agreement with the exact solution is obtained. As
the former calculation makes essential use of the nested
dipole hypothesis of Kosterlitz and Thouless, the validity of this
hypothesis is explicitly verified.
\end{quote}

\vspace{1.5cm}
\noindent
{\bf Key words:} Kosterlitz-Thouless transition; Coulomb gas;
renormalization equations; correlations; exact solution

\pagebreak

\normalsize
\noindent
{\bf 1.INTRODUCTION}
\vspace{5mm}

\noindent
The two-dimensional Coulomb gas refers to a neutral system of charged
particles confined to a plane. The two species have opposite charge
(magnitude $ q$ say) and interact via the laws of two-dimensional
electrostatics (logarithmic potential).To stop collapse between oppositely
charged particles at low temperature, due to the singular behaviour of the
logarithmic potential at the origin, a hard core or similar short range
regularisation is also required. For low densities and high temperature the
two-dimensional Coulomb gas forms a conductive phase in which the positive
and negative charges are dissociated and can screen a long-wavelength
external charge density.In contrast, for low densities  and low temperature,
the system forms a dipole phase in which the positive and negative particles
pair together . Perfect screening of a long-wavelength
external charge will no longer occur.

Intricate structures of nested dipoles were hypothesized by Kosterlitz and
Thouless [1] for the dominant configuration contributing to the
polarization as the transition point is approached from the dipole phase.
On the basis of this remarkable physical insight , an iterated mean- field
theory was formulated and quantities of physical interest thereby calculated
in the vicinity of the critical point.This so called Kosterlitz-Thouless
transition between the conductive and dielectric states occurs at the
coupling $\Gamma =4$ ($\Gamma :=q^{2}/kT$)in the zero-density limit.

The nested dipole hypothesis and the iterated mean-field equations of
Kosterlitz and Thouless were recently put on a firmer footing by Alastuey
and Cornu [2], who made a low fugacity ($\zeta$) analysis of the charge-charge
correlation function and the dielectric constant $\epsilon$ for $\Gamma
\rightarrow 4^+$. At order $\zeta^{4}$ it was proved that the configurations
giving the
leading order contribution to $1/\epsilon$ are the nested dipoles
hypothesized by Kosterlitz and Thouless. Assuming this to be true at all
orders, the low fugacity series could be resummed, and the iterated mean
field equations of Kosterlitz and Thouless derived exactly.

The pairing transition from a conductive to a dielectric phase is not unique
to the two-dimensional two-component Coulomb gas. One-component
log-potential Coulomb gases also exhibit this transition, provided
the neutralizing background consists of a lattice of oppositely charged
particles [3],or there is no background and the system is in the vicinity
of a conductive medium [4].
For the latter class of system a solvable model has been formulated which
exhibits a pairing transition as a microscopic parameter is varied [5].
The model is the  two-dimensional one-component plasma consisting of
particles of positive charge only confined to a half plane in the vicinity
of a metal wall, and subjected to a one-body external potential such that
$$
e^{-\beta V(y)}=y^{-\alpha}   \eqno(1.1)
$$
which acts in the direction perpendicular to the metal wall only. It is
solvable
at the special coupling $\Gamma =2$, and exhibits a pairing transition as
$\alpha$ is varied through one. It is our objective herein to use the
exact solution as a testbench for the predictions of theoretical methods
of analysing the Kosterlitz-Thouless transition for this model.

We begin in Section 2 by considering the screening properties of the system
with respect to an infinitesimal external dipole.
In Section 3 the low fugacity resummation technique of [2] is applied to
study the density profile in the dielectric phase near criticality, and
the density profile is further analysed using the first BGY equation. In
Section 4 comparison of the theoretical predictions with the exact results
are made. Concluding remarks are made in Section 5.

\vspace{1cm}
\noindent
{\bf 2.CHARACTERIZING THE PHASE}
\vspace{5mm}

\noindent
{\bf 2.1 Definition of the model}

\noindent
Consider a system of two-dimensional charges of strength $q$ confined to a
half plane $y\ge d$ and suppose a perfect conductor occupies the half plane
$y \le 0$. For each charge of strength $q$ at position $(x,y)$ say in the
system , the effect of the perfect conductor is to create an image charge
of strength $-q$ at position $(x,-y)$.The electrostatic potential
$\phi (\vec{r},\vec{r'})$ experienced by a test particle of charge $q$ at
$\vec{r'}=(x',y')$ due to a particle of charge $q$ at
$\vec{r}=(x,y)$ is then
$$
\phi (\vec{r},\vec{r'})=q^{2}\left( v_{c}(|\vec{r}
-\vec{r'}|)-v_{c}(|\vec{r}
-\vec{\bar{r'}})|\right)
\eqno (2.1a)
$$
where
$$
v_{c}(\vec{r}-\vec{r'})
=-\log[\{(x-x')^{2}+(y-y')^{2}\}^{1/2}/L]
\qquad\mbox{and}\qquad
\vec{\bar{r}}=(x,-y)
\eqno (2.1b)
$$

For convenience the arbitrary length scale $L$ will be henceforth set equal
to unity.
As well as interacting via the pair potential (2.1), the particles also
experience a one-body potential with Boltzmann factor (1.1). Since the
electrostatic potential $\phi (y) $ due to a background charge density
$q\rho_{b}(y^{'})$, $d\le y^{'}<\infty$ is given by
$$
\phi (y)=-\pi q\int_d^\infty dy'
\left( |y-y'|-(y+y')\right)
\rho_{b}(y')
\eqno (2.2)
$$
it is straightforward to check that the one-body potential given in (1.1)
can be interpreted as being due to a background charge density
$$
\rho_{b}(y')={\alpha \over 2\pi\Gamma {y'}^2}
\eqno (2.3)
$$
However it is more convenient for our purposes below to interpret (1.1) as
non-electrical in origin, and making no contribution to the total charge
density.
\vspace{5mm}

\noindent
{\bf 2.2 Response to an external dipole}

\noindent
The conductor and dipole phases of the two-dimensional Coulomb gas can be
distinguished by different screening properties of an infinitesimal
external charge: the external charge is perfectly screened in the conductor
phase
while only a fraction $1-1/\epsilon$
is screened in the dipole phase. For Coulomb systems near a metal
wall an external charge is automatically screened by its own image. We
consider instead the screening of an infinitesimal dipole. The image
of a dipole pointing perpendicular to a metal wall has the same magnitude
and direction as the original dipole. A conductor phase should perfectly
screen such an external dipole.

Let us use linear response theory to give a mathematical characterization
of the screening of an external dipole. An external dipole at
$\vec{r}_0:=(0,y_0)$ which is of strength $p_0$ and perpendicular to the metal
wall adds to the Hamiltonian a term
$$
H_{\rm ext}  =  p_0 \left[  \int_{\cal D} d\vec{r'} \,
{\partial \over \partial y_0} v_c(|\vec{r'}-\vec{r_0}|) Q(\vec{r'})
+ \int_{\cal D} d\vec{r'} \,
{\partial \over \partial \bar{y_0}} v_c(|\vec{r'}-\vec{\bar{r_0}}|) Q(\vec{r'})
\right] \eqno (2.4)
$$
where $\vec{\bar{r_0}}:=(0,-y_0),\bar{y_0}:=-y_0$  and $Q(\vec{r'})$
denotes the microscopic charge density at point $\vec{r'}$. The domain
${\cal D}$ is the half plane ${y \ge d}$. According to linear response theory
the change in charge density at a point $\vec{r}$, $\delta q(\vec{r})$ say,
due to the external dipole is given by
$$
\delta q(\vec{r})
= -\beta \left[ \langle H_{\rm ext} Q \rangle
- \langle H_{\rm ext}\rangle \langle Q \rangle \right]
\eqno (2.5)
$$
 From (2.4) the r.h.s. of (2.5) can be written in terms of the charge-charge
correlation
$$
S(\vec{r},\vec{r'}):= \langle Q(\vec{r})Q(\vec{r'}) \rangle
- \langle Q(\vec{r})\rangle \langle Q(\vec{r'})\rangle
\eqno (2.6)
$$
as
$$
\delta q(\vec{r}) = -\beta p_0 \int_{\cal D} d\vec{r'} \,
S(\vec{r},\vec{r'}) \left\{
{\partial \over \partial y_0} v_c(|\vec{r'}-\vec{r_0}|)+
{\partial \over \partial \bar{y_0}} v_c(|\vec{r'}-\vec{\bar{r_0}}|)
\right\}
\eqno (2.7)
$$
On the other hand, our characterization of the conductor phase as perfectly
screening the dipole says
$$
\int_{\cal D }d\vec{r} \, y\delta q(\vec{r}) = -p_0
\eqno (2.8)
$$
Substituting (2.7) in (2.8) gives the sum rule
$$
\beta \int_{\cal D} d\vec{r} \,y \int_{\cal D} d\vec{r'} \,
S(\vec{r},\vec{r'}) \left\{
{\partial \over \partial y_0} v_c(|\vec{r'}-\vec{r_0}|)+
{\partial \over \partial \bar{y_0}} v_c(|\vec{r'}-\vec{\bar{r_0}}|)
\right\}=1
\eqno (2.9)
$$
to be obeyed by the system in the conductor phase.

The sum rule (2.9) can be further simplified. Now
$$
 \int_{\cal D} d\vec{r} \,y \int_{\cal D} d\vec{r'} \,
S(\vec{r},\vec{r'})
{\partial \over \partial y_0} v_c(|\vec{r'}-\vec{r_0}|)
= \int_d^\infty dy \, y{\partial \over \partial y_0} F(y_0,y)
\eqno (2.10a)
$$
where
$$
F(y_0,y) := \int_{-\infty}^\infty dx \int_{-\infty}^\infty dx'
\int_d^\infty dy' S(y',y;x'-x)v_c(x'-x_0;y'-y_0)
\eqno (2.10b)
$$
{}From the convolution formula for Fourier transforms we have
$$
F(y_0,y) = \int_d^\infty dy' \, \tilde{S}(y',y;0)\tilde{v_c}
(0;y'-y_0)
\eqno (2.11)
$$
In performing this step we are assuming that with $y$ fixed
$S(x'-x;y',y)$ decays sufficiently as a function of $y'$ and $x-x'$
for the integral in (2.10b) to be absolutely
convergent and thus the order of integration to be unimportant.
Next, it is a straightforward exercise to deduce from
Poisson's equation
$$
{\nabla}^2 \, v_c(|\vec{r}|) = -2\pi \delta
(\vec{r})
\eqno (2.12)
$$
that
$$
\tilde{v_c}(0;y'-y_0) = -\pi |y'-y_0| \eqno (2.13)
$$
Substituting (2.13) in (2.11) and then substituting the resulting
expression in (2.10) and performing the differentiation gives
$$
 \int_{\cal D} d\vec{r} \,y \int_{\cal D} d\vec{r'} \,
S(\vec{r},\vec{r'})
{\partial \over \partial y_0} v_c(|\vec{r'}-\vec{r_0}|)
= \pi \int_d^\infty dy \, y \int_d^\infty dy' \,\tilde{S}(y',y;0)
{\rm sgn}(y'-y_0)
\eqno (2.14)
$$
Performing an analogous simplification on the second term on the
l.h.s. of (2.9) we thus deduce that (2.9) is equivalent, subject to the
 clustering assumption for $S(x'-x;y',y)$ noted below (2.11), to the
 simpler sum rule
$$
2\pi \beta \int_d^\infty dy \, y \int_{y_0}^\infty dy'
\int_{-\infty}^{\infty} dx' \; S(y',y;x'-x) =1
\eqno (2.15)
$$
This sum rule is to be satisfied in the conductive phase of any two-
dimensional Coulomb system separated a distance $d$ from a metal wall
(when $y_0=0$ this result has been obtained previously by Jancovici [6]).

A remarkable property of (2.15) is that it must hold for all positions $(0,
y_0)$ of the external dipole. Differentiating with respect to $y_0$, and
changing variables $x' \mapsto x'+x-x_0$, we thus have
$$
\int_{\cal D} d\vec{r} \, y S(\vec{r_0},\vec{r}) = 0.
\eqno (2.16)
$$
Hence the perfect screening of an external dipole implies the dipole
moment of the internal screening cloud must vanish.
For a phase which does not perfectly screen an external dipole, the
sum rule (2.15) cannot hold. Rather, we would expect the l.h.s. to depend on
$y_0$ and thus the dipole moment of the internal screening cloud will be
non-zero.

\vspace{5mm}
\noindent
{\bf 2.3 Phase transition and potential drop}

\noindent
In electrochemistry a fundamental quantity is the potential drop
across the interface:
$$
\Delta \phi = 2 \pi q \int_d^\infty dy \, y \rho(y)
\eqno (2.17)$$
where $q\rho (y)$ denotes the total charge density at distance $y$ from the
interface.
As previously noted [5], the formula (2.17) exhibits a further
interpretation of $\Delta \phi$: it is directly proportional to the mean
distance between a particle and the metal wall, or equivalently the mean
size of the particle-image pairs.
Therefore, $\Delta \phi$ is expected to be finite in the insulator phase,
while it should diverge in the conductor phase where the charges of the
plasma are not paired by their own images. As seen from the integral
expression (2.17), the finiteness of $\Delta \phi$ is closely related to
the large-distance behaviour of $\rho(y)$. In the insulator phase,
$\rho (y)$ should decay as $1/y^{2 + \epsilon}$ ($\epsilon > 0$) when
$y \rightarrow \infty$, while in the conductor phase $\rho (y)$ should decay
typically as $1 / y^2$ or slower. The relation between $\Delta \phi$ and
the asymptotics of $\rho (y)$ will be studied in Section 3, by resumming the
low fugacity expansions.

The potential drop $\Delta \phi$, or the internal dipole moment
$$
D(y_0) := 2 \pi \int_{\cal D} d\vec{r} \, yS(\vec{r_0},\vec{r})
\eqno (2.18)
$$
may be taken as equivalent idicators for characterizing the phase of the
present model. A relation between both quantities can be obtained by
starting with the compressibility sum rule
$$
\zeta  {\partial \rho(y) \over \partial \zeta}  =
{1 \over q^2} \int_{-\infty}^{\infty} dx' \int_d^\infty dy' \,  S(y,y';x-x')
\eqno (2.19)
$$
Taking the first moment of both sides gives
$$
\zeta \int_{d}^\infty dy \, y {\partial \rho(y) \over \partial \zeta}  =
{1 \over q^2}  \int_d^\infty dy \, y\int_{-\infty}^{\infty} dx'
\int_d^\infty dy'\,  S(y,y';x-x')
\eqno (2.20)
$$
Next we want to interchange the order of the $y$ and $y'$ integrations on
the r.h.s. and interchange the order of integration and differentiation on
the l.h.s. . From (2.17) and the explicit form of $S(y,y';x-x')$ (see Section
4)
we see that a necessary condition for the validity of both these
operations is that $\Delta \phi$ be finite.
Assuming this condition and the validity of the operations for the dipole
phase, we obtain from (2.20), after using (2.17) and (2.18), the desired
relationship:
$$
{\zeta {\partial \over \partial \zeta}  \Delta \phi}=
{1 \over q} \int_d^\infty dy' \, D(y')
\eqno (2.21)
$$

This equation can only be valid in the insulator phase, since in the conductor
phase,
the quantity on the r.h.s. of (2.20) is given by the universal value (2.15),
so we have instead
$$
\zeta \int_{d}^\infty dy \, y  {\partial \rho(y) \over \partial \zeta}  =
{1\over 2 \pi \Gamma}
\eqno (2.22)
$$
We stress that, in the conductor phase, the integrations over $y$ and $y'$
in the r.h.s. of (2.20) cannot be inverted. Otherwise, the corresponding
integral, which also appears in the l.h.s. of (2.15), would vanish as a
consequence of $D(y)=0$. This non-absolute convergence is related to a
slow decay of $S$ for some configurations. At the same time, the
differentiation with respect to $\zeta$ and the integration over $y$ in the
l.h.s. of (2.20) cannot be inverted either, because of a slow $1/y^2$-decay
of $\rho (y)$ (see Section 3).

\vspace{1cm}
\noindent
{\bf3.  THE DENSITY PROFILE}

\vspace{.5cm}
\noindent
A feature of the two-dimensional Coulomb gas is that all coefficients in the
low fugacity expansion of the pressure and correlation functions are
convergent for $\Gamma \ge 4$ [7]. This signals the transition from a
conductive phase for $\Gamma < 4$ to an insulator phase for $\Gamma \ge 4$,
in the zero-density limit.
Similarly, by examining the second moment
of the cluster integral for the density profile of a single particle, we
expect that all the  coefficients in the low fugacity expansions for the model
of subsection 2.1 are convergent for $\Gamma + 2\alpha > 4$ and that this
signals the transition from a conductive phase for
$\Gamma + 2 \alpha \le 4$ to an insulator phase for $\Gamma + 2 \alpha > 4$,
in the zero density limit.
In this
section the low fugacity expansion for the density profile will be studied
for $\Gamma + 2\alpha \rightarrow 4^+$, which is the limit of approaching
the phase boundary from the dipole side,
using the techniques introduced in
[2]. More precisely, we will study the asymptotic density profile
$\rho_{\Delta \phi}(y)$, which is defined as the portion of the low
fugacity expansion of $\rho(y)$ that gives the correct leading order
singular behaviour of each term in the low fugacity expansion of $\Delta
\phi$ (2.17) in this limit.

Alastuey and Cornu [2] complemented their study of the low fugacity
expansions of the correlations in the dipole phase of the two-dimensional
Coulomb gas by an analysis of the BGY equations. In subsection 3.5
the first BGY equation is used to compute the leading asymptotic behaviour
of the density profile. Unlike the resummation of the low fugacity
expansion calculation, there is no underlying assumption that the phase of the
model is
near the critical point on the dipole side.

\vspace{.5cm}
\noindent
{\bf 3.1 The expansion at $O(\zeta^2)$}

\noindent
Suppose the model of subsection 2.1 is generalized so that each particle
is associated with a position dependent fugacity $\zeta \mapsto \zeta
(y)$ (or equivalently is subject to an extra one body potential
acting perpendicular to the interface). Denote  the corresponding $N$-
particle canonical partition function by $Z_N$ and grand partition function by
$\Xi$. Then from the formula
$$
\rho (y) = \left. \zeta {\delta \over \delta \zeta(y)} \log \Xi \right |_{\zeta
(y)=\zeta}
\eqno (3.1)
$$
it is easy to show that
$$
\rho(y) = \left. \zeta {\delta \over \delta \zeta(y)} Z_1
\right|_{\zeta(y)=\zeta}
+\zeta ^2 \left (\left.  2 {\delta \over \delta \zeta(y)}Z_2
\right|_{\zeta(y)=\zeta}
-Z_1 \left. {\delta \over \delta \zeta(y)} Z_1 \right|_{\zeta(y) =\zeta} \right
) + O(\zeta ^3)
\eqno (3.2)
$$
Indeed (3.2) applies to any one-component system.  Inserting the form of the
partition functions for the model under consideration we have
$$
\rho(y) = {\zeta \over (2y)^{\Gamma /2} y^\alpha}
+ {\zeta^2  \over (2y)^{\Gamma /2} y^\alpha}
\int_{-\infty}^{\infty}dx_1 \, \int_d^\infty dy_1 \,{1 \over (2y_1)^{\Gamma
/2}{y_1}^\alpha}
$$
$$
\times
\left \{ \left ({(x-x_1)^2 + (y-y_1)^2 \over (x-x_1)^2
+ (y+y_1)^2 } \right ) ^{\Gamma /2} - 1 \right \}
+ O(\zeta ^3)
\eqno (3.3)
$$

Substituting the term of (3.3) proportional to $\zeta$, $\rho^{(1)}(y)$ say,
in (2.17) we find
$$
\Delta \phi^{(1)}/2 \pi q =
{\zeta 2^{-\Gamma /2} d^{-\Gamma /2 - \alpha + 2} \over \Gamma /2 - \alpha +2}
\eqno (3.4)
$$
(again, and below, we have used the superscript to  indicate that only the
term proportional to this power of $\zeta$ is being considered). Thus
$\Delta \phi^{(1)}$ is singular in the limit $\Gamma + 2 \alpha
\rightarrow 4^+$, and furthermore to leading order is independent of $d$.
Both features are true of $\Delta \phi^{(n)}$ in general.
 The latter feature implies that only the large-$y$ asymptotic
portion of $\rho^{(n)}(y)$ contributes to the leading order singular
behaviour of $\Delta \phi^{(n)}$, and thus $\rho_{\Delta \phi}^{(n)}(y)$
consists of terms in the asymptotic expansion of $\rho^{(n)}(y)$. With
$n=1$, there is only one term, which is $\rho^{(1)}(y)$ itself, so trivially
$\rho_{\Delta \phi}^{(1)}(y) = \rho^{(1)}(y)$.

Let us now determine  the portion of the term of order $\zeta ^2$ in the low
density
expansion  of $\rho (y)$ which contributes to the leading order singular
behavior of $\Delta \phi$, and thus calculate
$\rho _{\Delta \phi}^{(2)}(y)$.
For this purpose we consider the double
integral which is part of the coefficient of the $\zeta ^2$ term in (3.3)
and analyse it for
large-$y$.
We first break the integration over $y_1$ in the double integral in (3.3)
into two intervals: $[d, y]$ and $[ y,\infty]$. A change of variables
$y_1 \mapsto yv_1$ shows that the latter interval of integration gives a
contribution to the asymptotic expansion of ${\rho}^{(2)}(y)$ which is
$O(y^{-\Gamma /2 -\alpha -(\Gamma /2 + \alpha -2)})$, and a corresponding
contribution to $\Delta \phi ^{(2)}$ which is $O(1/(\Gamma + 2 \alpha - 4
))$. For the interval $[d, y]$, use of the large-$y$ expansion
$$
\left ( { (x-x_1)^2 + (y-y_1)^2 \over (x-x_1)^2 + (y+y_1)^2 } \right )
^{\Gamma /2} - 1
\: \sim \:
-{ 2\Gamma y_1 y \over x^2 + y^2}
\eqno (3.5)
$$
and integration over $x$ gives a contribution to the asymptotic expansion
of $\rho ^{(2)}(y)$ of
$$
{\zeta ^2 \over (2y)^{\Gamma/2}y^\alpha}\int_d^y dy_1 \, {\cal S}_y (y_1)
\eqno (3.6)
$$
where
$$
{\cal S}_y (y_1) := - {2 \pi \Gamma \over (2y_1)^{\Gamma /2} y_1^{\alpha - 1}},
\qquad d \le y_1 \le y
\eqno (3.7)
$$

Computing the integral, (3.6) reads
$$
-{2 \pi \Gamma \zeta ^2 \over 2^\Gamma y^{\Gamma /2 + \alpha}(\Gamma /2 +
\alpha -2)} \left ( d^{-(\Gamma /2 + \alpha - 2 ) } - y^{-(\Gamma /2 +
\alpha - 2 )} \right )
\eqno (3.8)
$$
Note that the first term in the last brackets above is all that need be
included for the leading order asymptotic expansion of $\rho ^{(2)} (y)$.
However both terms give a contribution to $\Delta \phi ^{(2)}$ which is
$O(1/(\Gamma + 2\alpha - 4)^2)$. Hence ${\rho_{\Delta \phi}^{(2)}(y)}$
is
given by both terms in (3.8), or equivalently the integral formula (3.6).
Following
[2] we can interpret the integral (3.6) as resulting from the partial
screening of the fixed particle-image pair of separation $2y$
by the smaller pair of separation $2y_1$, via the operator (3.7).

\vspace{.5cm}
\noindent
{\bf 3.2 Nested dipole hypothesis}

\noindent
Rather than attempt to calculate $\rho_{\Delta \phi }^{(n)}(y)$, $n \ge 3$,
from the low fugacity expansion (3.2), we make a nested dipole chain
hypothesis, analogous in idea to that of Kosterlitz and Thouless [1] and
technically to that of Alastuey and Cornu [2]. Technically  we suppose all
configurations contributing to $\rho_{\Delta \phi}^{(n)}(y)$ are
nested chains of particle- image pairs, with the fixed particle-image pair
the largest, and the screening operator acting between connected particles
in the chain only. To specify the chains we can ignore the images and
consider the different ways of arranging the mobile particles into chains
below the root particle at $y$.
For example, at $O(\zeta ^4)$ there are four distinct
configurations, which are illustrated graphically in fig. 1. The ordering
$y \ge y_1 \ge y_2 \ge y_3 \ge d$ is equivalent to the nesting of the
particle-image pairs so that each pair screens a pair of smaller size.
The
contribution to $\rho_{\Delta \phi}^{(4)}(y)$ from each graph is
$$
\left [ \int_d^y dy_1 \, {\cal S}_y (y_1) \int_d^{y_{1}} dy_2 {\cal S}_{y_1}
(y_2) \right ]
\left [ \int_d^y dy_1 \, {\cal S}_y (y_1) \right ],
\qquad
\left [ \int_d^y dy_1 \, {\cal S}_y (y_1) \right ]^3
$$
$$
 \int_d^y dy_1 \, {\cal S}_y (y_1) \left (\int_d^{y_1} dy_2 {\cal S}_{y_1}
(y_2)
 \right )^2
\qquad
 \int_d^y dy_1 \, {\cal S}_y (y_1) \int_d^{y_{1}} dy_2 {\cal S}_{y_{1}} (y_2)
\int_d^{y_{2}} d{y_3} {\cal S}_{y_2} ({y_3})
$$
respectively.
Furthermore, the graphs need to be weighted by factors of 6,1,3,6
respectively to account for relabelling degeneracy, and this linear
combination then multiplied by an overall factor of $\zeta^4  / 3!(2y)^
{\Gamma /2} y^{\alpha}$.

\vspace{1cm}

\setlength{\unitlength}{1cm}
\begin{minipage}[b]{4cm}
{}\hfill
\end{minipage}
\begin{minipage}[t]{13cm}
\begin{picture}(13,3)
\multiput(-1.5,3)(0,-1){4}{\line(1,0){.25}}
\put(-2,3){$y$}
\put(-2,2){$y_1$}
\put(-2,1){$y_2$}
\put(-2,0){$y_3$}
\multiput(.5,3)(2,0){4}
{\line(1,0){1}}
\thicklines
\multiput(1,3)(2,0){4}{\circle*{0.2}}
\put(1,3){\line(-1,-2){.5}} \put(1,3){\line(1,-2){.5}}
\put(3,3){\line(-1,-2){.5}} \put(3,3){\line(1,-2){.5}}
\multiput(3,3)(2,0){3}{\line(0,-1){1}}
\multiput(.5,2)(1,0){4}{\circle*{0.2}}
\multiput(3,2)(2,0){3}{\circle*{0.2}}
\put(.5,2){\line(1,-2){.5}}
\put(5,2){\line(-1,-2){.5}}\put(5,2){\line(1,-2){.5}}
\put(7,2){\line(0,-1){1}}
\multiput(1,1)(6,0){2}{\circle*{0.2}}
\multiput(4.5,1)(1,0){2}{\circle*{0.2}}
\put(7,1){\line(0,-1){1}}
\put(7,0){\circle*{0.2}}
\end{picture}
\end{minipage}

\vspace{.5cm}
\noindent
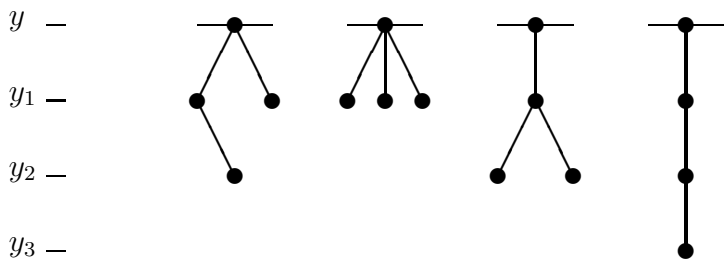
{\small {\bf Figure 1.} Graphical representation of the four distinct
chains at $O(\zeta^4)$}
\vspace{.5cm}

The nested structure allows the general terms of order $n$ in $\zeta$ to be
calculated by recurrence. As is shown in detail in [2, eqs.(4.26)-(4.28)]
we have
$$
\rho_{\Delta \phi}^{(n)}(y) = {\zeta \over (2y)^{\Gamma /2} y^\alpha}
{\zeta ^{n-1} \over (n-1)!} S_{\Delta \phi}^{(n-1)}(y)
\eqno (3.9a)
$$
where
$$
S_{\Delta \phi}^{(n-1)}(y)
= \sum_{p=1}^{n-1} { (n-1)! \over p! (n-1-p)!}
\sum_{q_{\alpha} \ge 0 \atop q_1+ \dots + q_p = n-1-p}
{(n-1-p)! \over q_1! \dots q_p!}I_{q_1}(y) \dots I_{q_p}(y)
\eqno (3.9b)
$$
with
$$
I_q(y) = \int_d^y dy' \, {\cal S}_y(y') S_{\Delta \phi}^{(q)}(y')
\eqno (3.9c)
$$
Furthermore, it is shown in [2] that $\rho_{\Delta \phi}^{(n-1)}(y)$ as
given by this equation can be summed over $n$ whatever the form of ${\cal
S}_y(y')$, with the result
$$
\rho_{\Delta \phi}(y) = {\zeta \over (2y)^{\Gamma /2} y^\alpha}
\exp \left [ \int_d^y dy_1 \, {\cal S }_y (y_1) (2y_1)^{\Gamma /2}
y_1^\alpha {\rho}_{\Delta \phi}(y_1) \right ]
\eqno (3.10)
$$
Inserting the explicit form (3.7) gives the non-linear integral equation
$$
{\rho_{\Delta \phi}}(y) = {\zeta \over (2y)^{\Gamma /2} y^\alpha}
\exp \left [ -2 \pi \Gamma \int_d^y dy_1 \, y_1 \rho_{\Delta \phi} (y_1)
\right ]
\eqno (3.11)
$$
which uniquely determines $\rho_{\Delta \phi}(y)$. Note that even though this
equation was derived with the assumption $\Gamma + 2\alpha \rightarrow 4^{+}$
at each order in $\zeta$,
it is well defined for all values of $\Gamma + 2 \alpha$.

\vspace{.5cm}
\noindent
{\bf 3.3 Evaluation of $\rho_{\Delta \phi}(y)$}

\noindent
By multiplying both sides of (3.11) by $y^{\Gamma /2 + \alpha}$ and
differentiating with respect to $y$ we obtain the non-linear differential
equation
$$
{d \over dy} g(y) = -{2 \pi \Gamma \over y^{\Gamma /2 + \alpha - 1}}
[g(y)]^2
\eqno (3.12a)
$$
where
$$
g(y) := y^{\Gamma /2 + \alpha} \rho_{\Delta \phi}(y)
\eqno (3.12b)
$$
This is to be solved subject to the initial condition $\rho_{\Delta \phi}(d)
=\zeta / [(2d)^{\Gamma /2}d^\alpha]$, obtained by substituting $y=d$ in
(3.11).Since the differential equation is first order separable, the
solution of the initial value problem is straightforward, and we find
$$
\rho_{\Delta  \phi}(y) = { \zeta / (2y)^{\Gamma /2} y^\alpha
 \over 1+ {2\pi \Gamma \zeta \over 2^{\Gamma /2}(\Gamma /2 + \alpha - 2)}
(d^{2- \Gamma /2 - \alpha} - y^{2 - \Gamma /2 - \alpha} ) }
\eqno (3.13a)
$$
provided $\Gamma + 2\alpha \ne 4$, while
$$
\rho_{\Delta  \phi}(y) = { \zeta / (2y)^{\Gamma /2} y^\alpha
 \over 1 + 2^{1- \Gamma /2} \pi \Gamma \zeta \log y }
\eqno (3.13b)
$$
for $ \Gamma + 2\alpha = 4$.

Inspection of (3.13a) show that the exponents in the y-dependent terms of
$\rho_{\Delta \phi}^{(n)}(y)$ depend on $\Gamma /2 + \alpha$ only. In
particular there is no dependence on the fugacity $\zeta$ and consequently
the phase transition will be independent of $\zeta$. This behaviour is to
contrasted with the two-dimensional Coulomb gas, in which the powers in the
decay of the asymptotic charge-charge correlation at order $\zeta^{(2n)}$
(the quantity analagous to $\rho_{\Delta \phi}(y)$) are dependent on $\zeta$
and the phase transition is $\zeta$-dependent.

We can use the resummations (3.13) to calculate $\Delta \phi$ to all  orders
in $\zeta$. First, from (2.17) we see that we require
$$
\rho_{\Delta \phi}(y) = O({1 \over y^{2+ \epsilon}}), \qquad \epsilon >0
$$
for $\Delta \phi$ to be finite. Now from (3.13) we have
$$
\rho_{\Delta \phi}(y) \sim
 \cases{
{ {2 -\Gamma /2 - \alpha  \over 2 \pi
\Gamma y^2 }},&$\Gamma + 2 \alpha < 4  $ \cr
{ {1 \over 2 \pi \Gamma y^2 \log y }},&$ \Gamma + 2 \alpha = 4$\cr
{ {c_{\zeta \Gamma \alpha} \over y^{\Gamma /2 + \alpha} }},&$\Gamma + 2
\alpha
>4  $  \cr}
\eqno (3.14)
$$
Hence $\Delta \phi$ is finite for $\Gamma + 2 \alpha > 4 $ (the dipole
phase), and infinite for $\Gamma + 2 \alpha \le 4$ (the conductor phase),
independent of the fugacity $\zeta$  as anticipated above.

The explicit value of $\Delta \phi$ in the insulator phase is given by
computing the integral (2.17) with $\rho (y)$ replaced by (3.13a). We find
$$
\Delta \phi /2 \pi q = {1 \over 2 \pi \Gamma} \log \left [ 1 + {
2 \pi \Gamma \zeta d^{2 - \Gamma /2 - \alpha} \over 2^{\Gamma /2}
(\Gamma /2 + \alpha - 2) } \right ]
\eqno (3.15)
$$
Again we emphasize that even though the intermediate steps leading to this
result require $ \Gamma + 2 \alpha \rightarrow 4^+$ at each order in
$\zeta$,
(3.15) is well defined for all $\Gamma + 2\alpha > 4$.

\vspace{.5cm}
\noindent
{\bf 3.4 Renormalization flow equation}

\noindent
In the two-dimensional Coulomb gas the renormalization flow equation relates
the asymptotic charge-charge correlation to the space dependent dielectric
constant, with the space variable an implicit parameter. In the present
model we can obtain a renormalization flow equation by relating
$\rho_{\Delta \phi}(y)$ to
$$
\Delta \phi (y) := 2 \pi q \int_d^y dy_1 \, y_1 \rho_{\Delta \phi}(y_1)
\eqno (3.16)
$$
To do this we differentiate (3.16) and use the definition (3.12b) to obtain
$$
{d \Delta \phi (y) \over dy} = 2 \pi q y^{1 - \Gamma /2 - \alpha} g(y)
\eqno (3.17)
$$
Dividing (3.12a) and (3.17) then gives the desired equation
$$
{dg(y) \over d \Delta \phi (y)} = -{\Gamma \over q}g(y)
\eqno (3.18)
$$
This flow equation is subject to the initial condition $g(d) =  2^
{- \Gamma /2} \zeta$ and $\Delta \phi (d) = 0$, and its exact solution is thus
$$
g(y) = 2^{-\Gamma /2}\zeta e^{-\Gamma \Delta \phi (y) /q}
\eqno (3.19)
$$

 From (3.15), for $\Gamma + 2 \alpha > 4$ the allowed values of $\Delta \phi
(y) $ is the finite interval $[0,\Delta \phi / 2 \pi q]$, while for
$\Gamma + 2 \alpha \le 4$ $\Delta \phi (y)$ takes on all values in
$[0,\infty[$. The flow diagram obtained from (3.19) thus has the appearance
sketched in figure 2.

\vspace{6.5cm}
\noindent
{\small {\bf Figure 2.} The flow diagram, where we have written
$x(y) := \Gamma \Delta (y) /q$. The different trajectories correspond to
different values of
$\Gamma + 2 \alpha $. The trajectories terminate for $\Gamma + 2\alpha >
4$.}

\vspace{.5cm}
\noindent
{\bf 3.5 The first BGY equation}

\noindent
In this subsection we will complement the above low fugacity resummation
study by an  asymptotic analysis of the first BGY equation.
Let us denote the force on a particle at $\vec {r_1}$ due to a
particle at $\vec {r_2}$ by $\vec {F_{21}}$, so that
$$
\vec{ F_{21}} = - \vec{\nabla}_1 \phi (\vec{ r_1},\vec {r_2})
\eqno (3.20)
$$
where $\phi (\vec{r_1},\vec{r_2})$ is given by (2.1a). Furthermore denote the
force on a particle at
$\vec{r_1}$ due to the self image particle and the one body potential by
$\vec {{F^{\rm im}_1}}$ and $\vec {{F^{\rm ext}_1}}$ respectively so that
$$
\vec {{F^{\rm im}_1}} = -{q^2 \over 2 y_1} \vec j \qquad {\rm and} \qquad
\beta \vec {{F^{\rm ext}_1}} = - {\alpha \over y_1} \vec{j}
\eqno (3.21)
$$
Then in terms of these forces the first BGY equation for the system is
$$
\vec{\nabla}_1 \rho (\vec{r_1}) =
\beta \vec{{F^{\rm ext}_1}} \rho (\vec{r_1}) +
\beta \vec{{F^{\rm im}_1}} \rho (\vec{r_1}) +
\beta \int_{\cal D} d\vec{r_2} \, \vec{F_{21}} \rho^{(2)}(\vec{r_1},\vec{r_2})
\eqno (3.22a)
$$

Let us consider the $y$-components of this equation for $y_1 \rightarrow
\infty$. We might expect that in this limit we can replace
$\rho^{(2)}({\vec r}_1,{\vec r}_2)$ in the final term in (3.22a) by
$\rho({\vec r}_1) \rho({\vec r}_2)$, which is equivalent to saying that if
we
write the final term in (3.22a) as
$$
\beta \int_{\cal D} d{\vec r}_2 \, {\vec F}_{21} \rho({\vec r}_1)
\rho({\vec r}_2) + \beta \int_{\cal D} d{\vec r}_2 \, {\vec F}_{21}
\rho^{(2)T}({\vec r}_1,{\vec r}_2)
\eqno (3.22b)
$$
the term involving $\rho^{(2)T}({\vec r}_1,{\vec r}_2)$ decays faster for
large-$y_1$ than the term involving $\rho({\vec r}_1) \rho({\vec r}_2)$(which
is a mean-field term),
plus the one body forces on the r.h.s. of (3.22a)
. In
the appendix this latter statement is proved subject to a mild clustering
assumption. Thus, neglecting the second term in (3.22b), we are left with
the mean-field type equation
$$
{\partial \over \partial y_1} \rho(y_1) =
\left [ -{\Gamma + 2 \alpha \over 2 y_1} +
\beta \int_{\cal D} d\vec{r_2} \, (F_{21})_y \rho (y_2) \right ] \rho(y_1)
\eqno (3.23)
$$
The integral over ${\cal D}$ in this equation can be simplified:
\begin{eqnarray*}
& & \int_{\cal D} d\vec{r_2} \,(F_{21})_y \rho (y_2)  \\
&=& -q^2 \int_{-\infty}^{\infty} dx_2 \int_d^{\infty} dy_2
\left [ {\partial \over \partial y_1}v_c(|\vec{r_1}-\vec{r_2}|)+
{\partial \over \partial \bar{y_1}}v_c(|\vec{\bar{r_1}}-\vec{r_2}|) \right ]
\rho (y_2) \\
&=& -q^2  \int_d^{\infty} dy_2
\left [ {\partial \over \partial y_1}\tilde{v_c}(0;y_1-y_2)-
{\partial \over \partial y_1}\tilde{v_c}(0;y_1+y_2) \right ]
\rho (y_2) \\
&=& -2 \pi q^2 \int_{y_1}^\infty dy_2 \, \rho (y_2)
\end{eqnarray*}
(this result can also be derived from Gauss's theorem in electrostatics).
The mean-field equation now reads
$$
{\partial \over \partial y_1} \rho(y_1) =
\left [ -{\Gamma + 2 \alpha \over 2 y_1} -
2 \pi \Gamma \int_{y_1}^\infty dy_2 \rho(y_2) \right ] \rho (y_1)
\eqno (3.24)
$$

To solve this equation for large-$y_1$ we seek a solution of the form
$$
\rho (y_1) \sim {c \over {y_1}^p}
\eqno (3.25)
$$
Substituting this in (3.24) gives
$$
-{pc \over y_1^{p+1}} \sim -{(\Gamma + 2\alpha )c \over 2y_1^{p+1}}
- {2\pi \Gamma c^2 \over (p-1) y_1^{2p-1}}
\eqno (3.26)
$$
For $p > 2$ the second term on the right hand side of (3.26) can be ignored
and we obtain a solution provided
$$
p = (\Gamma + 2\alpha )/2  \qquad ({\rm and \: thus \: } \Gamma + 2\alpha > 4)
\eqno (3.27a)
$$
For $p=2$, (3.25) is an exact solution of (3.24) provided
$$
c = { 4 - \Gamma - 2\alpha \over 4 \pi \Gamma } \qquad {\rm and}\qquad
\Gamma + 2 \alpha < 4
\eqno (3.27b)
$$
For $\Gamma + 2\alpha = 4$ we find, after equating the first two orders on
both sides of (3.24), that
$$
\rho (y_1) \sim {1 \over 2 \pi \Gamma {y_1}^2 \log y_1 }
\eqno (3.27c)
$$
is an asymptotic  solution.

We emphasize that the above analysis is asymptotically exact and
non-perturbative: the asymptotic formulas obtained hold for all values of
$\Gamma, \alpha$ and $\zeta$. As such, we can use these results to test
the predictions (3.14) for the leading asymptotics of $\rho (y)$ as
derived from $\rho_{\Delta \phi}(y)$. Surprisingly the results obtained
from $\rho_{\Delta \phi}(y)$ are in complete agreement with the
non-perturbative exact results, even though it has been assumed in the
derivation of $\rho_{\Delta \phi}(y)$ that the phase of the model is near
the zero-fugacity critical point on the dipole side.

\vspace{1cm}
\noindent
{\bf 4. Comparison with the solvable case}

\vspace{.5cm}
\noindent
{\bf 4.1 The phase}

\noindent
When $\Gamma = 2$ the model of subsection 2.2 is exactly solvable for all
$\alpha$ [5]. The exact expressions for the density profile and truncated
two particle distribution function are
$$
\rho (y) = 2 \pi \zeta y^{-\alpha}
\int_0^\infty dt \, {e^{-4 \pi y t} \over 1 + 2 \pi \zeta \int_d^\infty
dY Y^{-\alpha} e^{-4 \pi Y t}}
\eqno (4.1)
$$
and
$$
\rho^T ({\vec r}_1,{\vec r}_2) = -(2 \pi \zeta)^2 (y_1y_2)^{-\alpha} \left |
\int_0^\infty dt \, {e^{2 \pi i x t}e^{-2 \pi (y_1+y_2) t} \over 1 + 2 \pi
\zeta \int_d^\infty
dY Y^{-\alpha} e^{-4 \pi Y t}} \right |^2
\eqno (4.2)
$$
These expressions were used in [5] to show that for $\alpha \le 1$ the
dipole moment of the internal screening cloud $D(y_0)$ as defined by (2.18)
vanishes,
while for $\alpha > 1$ it is non-zero. This behaviour was interpreted as
indicating that the system exhibits a conductive phase for $\alpha \le 1$
and an insulator phase for $\alpha  > 1$. In subsection 2.2 we have showed
that the true indicator of a conductive phase is the sum rule (2.15), and
the vanishing of $D(y_0)$ in a conductive phase is a corollary of this
stronger requirement.

Noting that for a one-component system
$$
S(y,y';x-x') = q^2 \left[ \rho (y') \delta (x-x') \delta (y-y')
+ \rho^T(y,y';x-x') \right ]
\eqno (4.3)
$$
and using the exact results (4.1) and (4.2), it is a straightforward
exercise to show
$$
2 \pi \beta \int_d^\infty dy \, y \int_{y_0}^\infty dy' \int_{-\infty}^
\infty dx' \, S(y,y';x-x')
= 1-
{ 1 + 2 \pi \zeta \int_d^{y_0} dY \, Y^{-\alpha} \over
 1 + 2 \pi \zeta \int_d^{\infty} dY \, Y^{-\alpha} }
\eqno (4.4)
$$
For $\alpha \le 1$ the second term on the r.h.s. of (4.4)
vanishes and the sum rule (2.15) holds, thus implying a conductive phase.
For $\alpha > 1$ (2.15) is not obeyed so the phase in an insulator. These
conclusions are in agreement with those reached in [5].

In subsection 3.3 we have shown that the potential drop $\Delta \phi $
diverges for $\Gamma + 2\alpha \le 4$ but is finite for $\Gamma + 2\alpha
> 4$. In accordance with the interpretation of the formula (2.17) as
saying
$\Delta \phi$ is  proportional to the mean distance of separation within
the dipole formed by a particle and its image, we have taken this behaviour
to be an alternative phase indicator to the sum rule (2.15) for this
system. For the solvable model we have the exact result [5]
$$
\Delta \phi / 2 \pi q = {1 \over 4 \pi}
\log \left ( 1+ 2 \pi \zeta {d^{1 - \alpha} \over \alpha - 1} \right ),
\qquad \alpha > 1
\eqno (4.5)
$$
Remarkably, the expression for $\Delta \phi$ (3.15) with $\Gamma = 2$
deduced from the asymptotic density profile $\rho_{\Delta \phi} (y)$ is in
precise agreement with this exact expression.

\vspace{.5cm}
\noindent
{\bf 4.2 The asymptotic density $\rho_{\Delta \phi}(y)$}

\noindent
The asymptotic density $\rho_{\Delta \phi}(y)$ is defined as the portion of
the asymptotic expansion of $\rho (y)$ that gives the correct singular
behaviour of $\Delta \phi$ as $\Gamma + 2\alpha \rightarrow 4^{+}$ at each
order in $\zeta$. From (4.1) we can calculate $\rho_{\Delta \phi}(y)$
exactly at $\Gamma = 2$.

Expanding (4.1) as a power series in $\zeta$ and then performing the
integration over $t$ gives
$$
\rho (y) = {\zeta \over 2y^{\alpha + 1}}
\sum_{j=0}^{\infty} (-2 \pi \zeta)^j y^{-j(\alpha + 1)}
\int_{[d/y,\infty]^j} {dY_1 \over {Y_1}^{\alpha}} \dots
 {dY_j \over{Y_j}^{\alpha}}
(Y_1 + \dots + Y_j + 1)^{-1}
\eqno (4.6)
$$
For large-$y$ the final factor in the integral can be approximated by 1 and we
obtain
$$
\rho (y) \sim {\zeta \over 2y^{\alpha + 1}}
\sum_{j=0}^\infty (-2 \pi \zeta)^j y^{-j(\alpha + 1)}
\left [ \left ( {(y/d)^{\alpha - 1} - 1) \over \alpha - 1} \right )^j
+ O([\alpha - 1]^{-j+1}[y/d]^{j(\alpha -1)-1}) \right ]
\eqno (4.7)
$$
The correction term in the asymptotic expansion above does not contribute to
$\rho_{\Delta \phi}(y)$. Ignoring this term, we see that a geometric series
remains, which after summation gives
$$
\rho_{\Delta \phi} (y) =  {\zeta / 2y^{\alpha + 1}
 \over 1 + 2 \pi {\zeta \over \alpha - 1} (d^{1 - \alpha} - y^{1 -
\alpha}) }
\eqno (4.8)
$$
Comparison of this exact result at $\Gamma = 2$ with the result (3.13a)
obtained from the low fugacity resummation using the nested dipole chain
hypothesis shows that the resummation is
exact at this coupling.

\vspace{.5cm}
\noindent
{\bf 4.3 Leading asymptotics of density profile}

\noindent
As noted in [5], the leading large-$y$ behaviour of the density profile at
$\Gamma = 2$ is readily computed from (4.1). We find precise agreement with
the behaviour (3.14) at $\Gamma = 2 $, which is obtained from both the low
fugacity resummation and the mean-field equation.

It is interesting to note that since the asymptotic form of the density profile
in the conductive phase is
$$
\rho (y) \sim {4 - \Gamma -2 \alpha \over 4 \pi \Gamma y^2 }
\eqno (4.9)
$$
the phase
transition occurs when the $1/y^2$ tail vanishes.

\vspace{1cm}
\noindent
{\bf 5. CONCLUSION}

\vspace{.5cm}
\noindent
The metal wall one-component plasma model of subsection 2.1 exhibits both
a conductive and insulating phase. It has the special property of admitting
an exact solution for the thermodynamics and all correlations along the line
$(\Gamma ,\alpha ) = (2,\alpha)$ in parameter space [5]. This line
intersects the transition line $\Gamma + 2\alpha =4 $. For general values of
the parameters the transition can be analysed in a similar way to the
Kosterlitz-Thouless transition in the two-dimensional Coulomb gas [2]. In
particular, by making an hypothesis that the dominant configurations are
nested dipole chains (this is checked explicitly at $O(\zeta^2)$), the low
fugacity expansion of the asymptotic density $\rho_{\Delta \phi} (y)$ can be
resummed, and calculated explicitly as the solution of a non-linear
differential equation.

Comparison with exact solution verifies that the general expression for
$\rho_{\Delta \phi}(y)$ is exact at $\Gamma = 2$. This provides compelling
evidence for the correctness of the underlying nested dipole chain
hypothesis. Since the nested dipole chain hypothesis also underlies the
iterated mean-field equations of Kosterlitz and Thouless [1] (which are
equivalent to the Kosterlitz renormalization equations [8]), we have also
added further weight to the validity of these equations.

\vspace{1cm}
\noindent
{\bf ACKNOWLEDGEMENT}

\vspace{.5cm}
\noindent
Most of this work was done during a visit by A. Alastuey to La Trobe
University during November and December 1993. Financial support for this
trip was provided by La Trobe and the CNRS. The research of P.J. Forrester
is supported by the Australian Research Council. Also, we thank Sam Yang
for generating figure 2.

\pagebreak

\noindent
{\bf Appendix}

\vspace{.5cm}
\noindent

In this appendix, we will prove that in the limit
$y_1 \rightarrow \infty$ the second term in (3.22b),
$$
{\vec{\cal F}}_1^{(2)} := \beta \int_{\cal D} d{\vec r}_2 \,
{\vec F}_{21}\rho_{(2)}^T
({\vec r}_1,{\vec r}_2)
\eqno ({\rm A}1)
$$
decays faster than the sum of the first term and the one body terms on the
r.h.s.  of the BGY equation (3.22a), and therefore can be neglected in this
limit. Our analysis is based on the simple assumption that for some
$1 > \epsilon > 0$
$$
|\rho_{(2)}^T({\vec r}_1,{\vec r}_2)| <
\rho({\vec r}_1) \rho({\vec r}_2)\left ( {l \over |{\vec r}_1 - {\vec r}_2|
} \right )^\epsilon
\eqno ({\rm A}2)
$$
where $l$ is a given length. We stress that the hypothesis (A2) is very
reasonable since it merely asserts that the Ursell function
$\rho_{(2)}^T({\vec r}_1,{\vec r}_2)/[\rho({\vec r}_1 \rho({\vec r}_2)]$
decays for large separations $|{\vec r}_1 - {\vec r}_2|$ at least as
fast as an inverse power. This weak clustering property surely holds in any
homogeneous or inhomogeneous fluid phase.

Using (A2) and the inequality $|y_2-y_1| \le |{\vec r}_2 -{\vec r}_1|$, we
find
\begin{eqnarray*}\lefteqn{
|{\vec{\cal F}}_1^{(2)}|}\\
& & <2 \Gamma \rho(y_1) \int_{\cal D} d{\vec r}_2 \,\rho(y_2)
\left ( {l \over |y_2 -y_1 | }\right )^\epsilon y_2
{ 1 \over [(x_2 -x_1)^2 + (y_2 - y_1)^2]^{1/2}[(x_2 - x_1)^2 + (y_2 +
y_1)^2]^{1/2}}
\end{eqnarray*}
$$
\eqno ({\rm A}3)
$$
In the integral on the r.h.s. of (A3), we can perform the integration over
$x_2$ according to [9]
$$
\int_{-\infty}^\infty dx_2 \,
{ 1 \over [(x_2 -x_1)^2 + (y_2 - y_1)^2]^{1/2}[(x_2 - x_1)^2 + (y_2 +
y_1)^2]^{1/2}}
= { 2 \over (y_2 + y_1) } K\left ( {2 \sqrt{y_2y_1} \over y_2 + y_1} \right
)
\eqno ({\rm A}4)
$$
where $K(k)$ is the complete elliptic integral of the first kind
$$
K(k) = \int_0^{\pi /2} {d \phi \over \sqrt{1 - k^2 \sin ^2 \phi}}
$$
By splitting the domain of integration over $y_2$ into
the intervals $[d,y_1/2]$ and $[y_1/2, \infty[$, we then find from (A3)
$$
|\vec{{\cal F}}_1^{(2)}| < 4 \Gamma K \left ({2 {\sqrt 2} \over 3} \right )
\left ( {2l  \over y_1}\right )^\epsilon {\rho(y_1) \over y_1}
\int_d ^{y_1/2} dy_2 \, y_2 \rho(y_2) \qquad { }
$$
$$
+ 4 \Gamma \rho(y_1) \int_{y_1/2}^\infty dy_2 \, \rho(y_2)
\left ( {l \over |y_2 - y_1| }\right )^ \epsilon
K \left ( 2\sqrt{y_2y_1} \over y_2 + y_1 \right )
\eqno ({\rm A}5)
$$
(we have also used the monotonicity of $K(k)$).

For $y_1$ large, $\rho(y_1)$ is expected to decay as $c /y_1^p$, with
$p \ge 2$ (this is shown explicitly in subection 3.5). The integral
$$
\int_d^{y_1/2} dy_2 \, y_2 \rho (y_2)
$$
then remains bounded by some constant times $\log (y_1)$.
Therefore the first term on the r.h.s. of (A5) decays at least as fast as
$\log (y_1) / y_1^{p+1+\epsilon}$ which is faster than the one-body self image
and
external potential terms
($\sim 1/y_1^{p+1}$) appearing on the r.h.s. of the BGY equation (3.22a). Also,
the integral
$$
\int_{y_1/2}^\infty dy_2 \, \rho(y_2) \left ({l \over |y_2 - y_1|}
\right )^\epsilon K \left ( {2 \sqrt{y_2y_1} \over y_2 + y_1 } \right )
$$
remains bounded by a constant times $1/y_1^{p-1+\epsilon}$, as shown by the
variable change $y_2 = \alpha y_1$. Indeed, the dimensionless integral
$$
\int_{1/2}^\infty d\alpha {1 \over \alpha^p |\alpha - 1|^\epsilon}
K\left ( {2 \alpha^{1/2} \over \alpha + 1} \right )
$$
is finite because the singularity of $K(2 \alpha^{1/2}/(\alpha + 1))$
at $\alpha = 1$ is only logarithmic:
$$
K(2 \alpha^{1/2}/(\alpha + 1)) \: \sim \: - \log |\alpha - 1|
$$
when $\alpha \rightarrow 1$. Then, the second term on the r.h.s. of (A5)
decays at least as fast as $1/y_1^{2p - 1 +\epsilon}$, which is faster than
the decay of the first term in (3.22b) $(\sim 1 / y_1^{2p - 1})$. Thus the
whole two-body force (A1) can be neglected with respect to the other  terms
of the BGY equation (3.22a) in the limit $y_1 \rightarrow \infty$.

\pagebreak

\noindent
{\bf REFERENCES}

\vspace{.5cm}
\noindent
[1] J.M. Kosterlitz and D.J. Thouless, J. Phys. C 6 (1973) 1181 \\[.2cm]
[2] A. Alastuey and F. Cornu, J. Stat. Phys. 66 (1992) 165 \\[.2cm]
[3] J. Cl\'{e}rouin and J.P. Hansen, Phys. Rev. Lett. 54 (1985) 2277;
A. Alastuey, F. Cornu and B. Jancovici, Phys. Rev. A 38 (1988) 4916 \\
[.2cm]
[4] A. Alastuey, Mol. Phys. 52 (1984) 637 \\[.2cm]
[5] P.J. Forrester, Int. J. Mod. Phys. A 7, suppl. 1A, (1992) 303 \\[.2cm]
[6] B. Jancovici, J. Phys. (Paris) 47 (1986) 389; Ph. A. Martin, Rev. Mod.
Phys. 60 (1988) 1075 \\[.2cm]
[7] E.R. Speer, J. Stat. Phys. 42 (1986) 895 \\[.2cm]
[8] A.P. Young, J. Phys. C 11 (1978) L453 \\[.2cm]
[9] I.M. Ryshik and I.S. Gradstein, "Tables of Series, Products and
Integrals" (VEB Deutscher Verlag der Wissenschaften, Berlin, 1963)

\end{document}